\documentclass[twocolumn,showpacs,preprintnumbers,amsmath,amssymb]{revtex4}
\usepackage[dvips]{graphicx}
\usepackage{dcolumn}

\begin{document}

\title
{Instability of square vortex lattice in $d$-wave superconductors is due to paramagnetic depairing
}

\author{Norihito Hiasa and Ryusuke Ikeda}

\affiliation{%
Department of Physics, Kyoto University, Kyoto 606-8502, Japan
}

\date{\today}


\begin{abstract} 
Effects of the paramagnetic depairing on structural transitions between vortex lattices of a quasi two dimensional $d$-wave superconductor are examined. It is found that, in systems with Maki parameter $\alpha_{\rm M}$ of order unity, a square lattice induced by a $d$-wave pairing is destabilized with {\it increasing} fields, and that a reentrant rhombic lattice occurs in higher fields. Further, a weak Fermi surface anisotropy competitive with the pairing symmetry induces another structural transition near $H_{c2}$. These results are consistent with the structure changes of the vortex lattice in CeCoIn$_5$ in ${\bf H} \parallel c$ determined from recent neutron scattering data. 
\end{abstract}



\maketitle

Recently, various novel superconducting (SC) properties have been found in the heavy fermion superconductor CeCoIn$_5$ in magnetic fields. Among them, the discontinuous $H_{c2}$-transition \cite{Izawa} and a new high field phase, identified with a Fulde-Ferrell-Larkin-Ovchinnikov (FFLO) vortex state \cite{Bianchi03,MS,AI,RI07}, have been the subjects of central interest in the research field. Quite recently, much attention has been paid, in turn, to the vortex lattice in fields (${\bf H} \parallel c$) perpendicular to the SC layers. An unusual $H$ dependence of the structural factor of the lattice has been noticed and argued \cite{Bianchi08,Ocha} to be due to an interplay of the paramagnetic depairing \cite{Okayama} and a non-SC (magnetic) critical fluctuation, both of which become active on approaching $H_{c2}(0)$ from below. Another feature stressed there \cite{Bianchi08,Ocha} is a striking reentry of the region of a rhombic lattice which is accompanied by an instability of a square lattice induced by a $d$-wave pairing \cite{Affleck,Nakai} or a four-fold anisotropy of the Fermi surface (FS) \cite{Nakai}: With {\it increasing} $H$ in $H \geq 0.6 H_{c2}(0)$, the square lattice changes into a rhombic one. It will be valuable to clarify whether the origin of this reentry is due to the paramagnetic depairing, the non-SC critical fluctuation, or others \cite{Nakai,Kogan,Dorsey}. 

Below, we study changes of the vortex lattice structure induced by the Pauli paramagnetic depairing in a superconductor with a $d$-wave pairing. In the orbital limit with vanishing Maki parameter $\alpha_{\rm M}=\sqrt{2}H_{\rm orb}(0)/H_P(0)$, the familiar enhancement of square lattice symmetry in higher magnetic fields and upon cooling is obtained, where $H_{\rm orb}(0)$ is the orbital limiting field in 2D case, and $H_P(0)$ is the Pauli-limiting field. By contrast, our calculation using a Maki parameter $\alpha_{\rm M} > 2.5$ shows a phase diagram similar to the observed one in CeCoIn$_5$ \cite{Bianchi08}, where a reentry of the rhombic lattice region occurs with increasing $H$, and suggests that effects of the pairing symmetry on the structure and orientation of the vortex lattice are weakened with increasing the field by the paramagnetic depairing. In fact, a FS anisotropy competitive with the $d$-wave symmetry is found to, in higher fields, can change the orientation of the vortex lattice discontinuously. 
 The present results strongly suggest that the main origin of the square to rhombic reentrant structural transition (ST) curve seen in CeCoIn$_5$ is the paramagnetic depairing. In relation to this, the issue of the pairing symmetry of CeCoIn$_5$ will also be discussed. 

The theoretical method we use here is a straightforward extension of that in Ref.\cite{AI} starting from the 2D BCS hamiltonian with a circular FS
\begin{eqnarray}
{\cal H} &=& d \sum_{\sigma = \pm 1} \biggl[ \int d^2 r ( \, {\varphi}_{\sigma}({\bf r}) \, )^\dagger 
\Bigg[ \, \frac{ ({\rm -i}{\nabla} + e{\bf A} )^2}{2m} - \sigma \mu H \, \Bigg]  {\varphi}_{\sigma}({\bf r}) \nonumber \\
&-& \frac{|g|}{2} \int \frac{d^2 k} {(2 \pi)^2} B_{\sigma}^{\dagger}({\bf k}) B_{\sigma}({\bf k}) \biggr],
\label{BCSint}
\end{eqnarray}
to derive an appropriate Ginzburg-Landau (GL) free energy density ${\cal F}$ , where $\varphi_\sigma({\bf r}) = S^{-1/2} \sum_{\bf p} c_\sigma({\bf p}) e^{{\rm i}{\bf p}\cdot{\bf r}}$, $S$ is the system area, and $B_{\sigma}({\bf k}) = \sum_{\bf p} {\hat \Delta}_{\bf p} c_{-\sigma}({\bf -p_-}) \, c_{\sigma}({\bf p_+})$, where ${\bf p}_\pm = {\bf p \pm {\bf k}}/2$. Effects of including a FS anisotropy will be explained later. The normalized pairing function ${\hat \Delta}_{\bf p}$ will be assumed hereafter to be $\sqrt{2} \, {\rm cos}(2\phi_p)$, where $\phi_p={\rm arctan}(p_y/p_x)$. In ${\bf H} \parallel c$ of interest in this work, spatial variations parallel to ${\bf H}$ are, in the mean field theory, negligible in the equilibrium states in lower fields than the FFLO region \cite{MS,RI07} just below $H_{c2}$ . For this reason, we can focus hereafter on the 2D model (1) to discuss CeCoIn$_5$. 

In deriving ${\cal F}$ in ${\bf H} \parallel c$ from the model (1), two mixings neglected in Ref.\cite{AI} need to be incorporated here: First, in expressing the SC order parameter $\Delta$ in terms of the Landau level (LL) modes, a coupling or mixing, induced by the $d$-wave symmetry, between the lowest ($n=0$) and higher ($n = 4m \geq 4$) LLs will be incorporated in the GL term ${\cal F}_2$ quadratic in $\Delta$ and $\Delta^*$ because we are interested here in properties in intermediate magnetic fields rather than those close to $H_{c2}$ \cite{AI}. For simplicity, just the most dominant contribution among the higher LLs, the $n=4$ LL, will be kept below. Further, a small mixing, which occurs in the higher order GL terms ${\cal F}_m$ ($m \geq 4$), in the momentum space between the relative momenta of Cooper pairs and the reciprocal lattice vectors of $\Delta$ will be taken into account. The contribution to the sign and magnitude of ${\cal F}_m$ of the latter mixing is quite small and, in fact, was neglected in the previous work \cite{AI} where the global phase diagram was studied. However, it plays essential roles, together with the higher LL corrections, in studying stable lattice structures. 
\begin{figure}[t]
\scalebox{0.95}[0.95]{\includegraphics{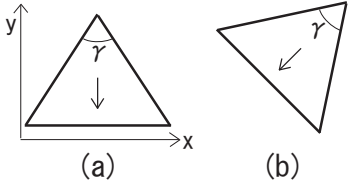}}
\caption{One half of a unit cell of a vortex lattice in real space oriented (a) by the $d_{x^2-y^2}$-pairing function ${\hat \Delta}_p$ or (b) by a four-fold FS anisotropy competitive with ${\hat \Delta}_p$. A rhombic to square ST is driven by the squashing indicated by each solid arrow.} \label{fig.1}
\end{figure}

Then, $\Delta$ is expressed as $\Delta({\bf r})=a_0 \, \varphi_0(x,y)+ a_4 \, \varphi_4(x,y)$ satisfying $\langle |\Delta|^2 \rangle_s=1$, where, in the Landau gauge ${\bf A}=Hx {\hat y}$, $\varphi_n=(n!)^{-1/2}(r_H(-{\rm i}\partial_y + 2eHx-\partial_x)/\sqrt{2})^n \varphi_0(x,y)$, $\varphi_0(x,y) = C_0 \sum_n \exp[{\rm i}(k \, n \, (y/r_H) + \pi n^2/2 ) - ( k n + x/r_H)^2/2 ]$ is the Abrikosov lattice solution in the $n=0$ LL with the orientation of Fig.1(a), $r_H^{-1}=\sqrt{2eH}$, and $\langle \,\,\,\rangle_s$ denotes spatial average. The order parameter of the square to rhombic ST is $k^2 - \pi$, where $k$ ($>0$) is defined by $
k^2 = \pi \, {\rm cot}(\gamma/2)$ 
in terms of the apex angle $\gamma$ (see Fig.1(a)). 

The quadratic GL term ${\cal F}_2$ takes the form 
\begin{equation}
{\cal F}_2 = N(0) [ {\cal M}_{00} |a_0|^2 + {\cal M}_{44} |a_4|^2 + {\cal M}_{0 4} (a_0^* a_4 + a_4^* a_0) ], 
\end{equation}
where 
\begin{eqnarray}
{\cal M}_{n_1 \, n_2} &=& \frac{1}{N(0)|g|} \, \delta_{n_1, n_2}  
- \int_0^\infty d\rho_0 \, f(\rho_0) \nonumber \\ 
&\times& \int_{-\pi}^{\pi} \frac{d\phi_p}{2 \pi} |{\hat \Delta}_p|^2 {\cal L}_{n_1  n_2}(\mu_0) \, \exp(-|\mu_0|^2/2), 
\end{eqnarray} 
${\cal L}_{n n}(\mu)=L_n(|\mu|^2)$ is the $n$th order Laguerre polynomial, ${\cal L}_{4 0}(\mu)=({\cal L}_{0 4}(\mu))^*=\mu^4/\sqrt{24}$, 
$\mu_j=\rho_j (v_x-{\rm i}v_y)/(\sqrt{2} r_H)$, 
\begin{equation}
f(\rho) = 2 \pi T \exp\biggl(- 2 \pi T_{c0} \, \rho \, \frac{\xi_0}{l} \biggr) \, \frac{{\rm cos}(2\mu_B H \rho)}{{\rm sinh}(2 \pi T \rho)}, 
\end{equation}
$l/\xi_0$ is a mean free path in the normal state normalized by the coherence length , and $T_{c0}$ is the (mean field) SC transition temperature in the case with $H=0$ and $l=\infty$. Further, any spatial variation of the flux density was neglected. 
After diagonalizing eq.(2), the $H_{c2}(T)$-curve is given by ${\cal M}_{00} {\cal M}_{44}-({\cal M}_{04})^2=0$. The eigenmode determining $H_{c2}(T)$ takes the form $\Delta={\rm cos}\chi \, \varphi_0 + {\rm sin}\chi \, \varphi_4$, where $\chi > 0$, and ${\rm cos}2\chi = ({\cal M}_{44} - {\cal M}_{00})/{\sqrt{({\cal M}_{44}-{\cal M}_{00})^2 + 4 {\cal M}_{04}^2}}$. 

Next, let us turn to the analysis of the quartic GL term ${\cal F}_4$, or equivalently, the Abrikosov factor $\beta_A$ determining the lattice structure. Hereafter, we use the gauge ${\bf A}=Hx{\hat y}$. To express ${\cal F}_4$ in a convenient form for numerical analysis, we use the relation 
\begin{eqnarray}
\Delta^{(n)}({\bf r};\mu) &\equiv& \exp({\rm i}\rho {\bf v}\cdot{(-{\rm i}\nabla+2e{\bf A})}) \, \varphi_n(x,y) \nonumber \\
&=& \frac{\exp(-|\mu|^2/2)}{\sqrt{n!}} \biggl(\mu^* \! - \! \frac{\partial}{\partial \mu} \biggr)^n  [ \, \exp(\mu^2/2) \nonumber \\
&\times& \varphi_0(x+\sqrt{2}r_H \mu, y) \, ], 
\end{eqnarray}
which can be obtained in the present gauge by extending the analysis in Ref.\cite{AI} to the case with higher LLs. The function $\varphi_n(x,y)$ itself is given by taking the $\mu$, $\mu^* \to 0$ limit in eq.(5). Then, ${\cal F}_4$ is expressed by \cite{AI} 
\begin{equation}
\frac{{\cal F}_4}{2 N(0)} = \int_0^\infty d\rho_1 d\rho_2 d\rho_3 \, f\biggl(\sum_{j=1}^3 \rho_j \biggr) \, {\cal J}_4, 
\end{equation}
where
\begin{eqnarray}
{\cal J}_4 &=& S^{-1} \int d^2{\bf r} 
\, \int_{-\pi}^{\pi} \frac{d\phi_p}{2 \pi} \, {\rm Re} \, [ \, |{\hat \Delta}_p|^4 \, \Delta({\bf r}) (\Delta({\bf r};-\mu_1))^* \nonumber \\
&\times& (\Delta({\bf r};-\mu_3))^*  \, \Delta({\bf r};\mu_2) \, ],  
\end{eqnarray}
and 
\begin{equation}
\Delta({\bf r};\mu_j) = {\rm cos}\chi \, \Delta^{(0)}({\bf r};\mu_j) + {\rm sin}\chi \, \Delta^{(4)}({\bf r};\mu_j).
\end{equation}

Below, other higher order GL terms will not be considered. This assumption is not permitted if ${\cal F}_4 < 0$. We will neglect the narrow region close to $H_{c2}(0)$ including the FFLO region \cite{AI,RI07} and focus on the field and temperature range with a positive ${\cal F}_4$. Then, by carrying out the ${\bf r}$-integral and the $\mu_j$-derivatives in eq.(6), ${\cal J}_4$ is given by 
\begin{eqnarray}
{\cal J}_4 &=& \frac{k \, {\rm cos}^4\chi}{\sqrt{2 \pi}}\sum_{m,n} (-1)^{mn} \int_{-\pi}^{\pi} \frac{d\phi_p}{2 \pi} \, |{\hat \Delta}_p|^4 \exp\biggl(-\frac{k^2(m^2+n^2)}{2} \nonumber \\
&-& \frac{1}{2} \sum_{j=1}^3 |\mu_j|^2 \biggr)  \, {\rm Re} \, [ \, e^{-p_0} ( 1 + p_1 \, {\rm tan}\chi 
+ p_2 \, {\rm tan}^2\chi  ) ],
\end{eqnarray}
up to O(${\rm tan}^2\chi$), where 
\begin{eqnarray}
p_0 &=& \frac{1}{2}(\mu_2^2 + (\mu_1^*)^2 + (\mu_3^*)^2) - \frac{1}{4} (\mu_2-\mu^*_1-\mu^*_3)^2 \nonumber \\
&-& \frac{k}{\sqrt{2}} (n(\mu_2+\mu_1^*-\mu_3^*)+m(\mu_2-\mu_1^*+\mu_3^*)), \nonumber \\
p_1 &=& \frac{1}{\sqrt{4!}} \sum_{j=1}^4 \biggl(\frac{3}{4} - 3K_j^2+K_j^4 \biggr), \nonumber \\
p_2 &=& \frac{1}{4!} \biggl( 9 + 2\sum_{i<j} [(-1)^{i+j} K_i K_j \biggl[ 6 + (3-2K_i^2)(3-2K_j^2) ] \nonumber \\
&+& \biggl(\frac{3}{4} - 3K_i^2+K_i^4 \biggr) \biggl(\frac{3}{4} - 3K_j^2+K_j^4 \biggr) \nonumber \\ 
&+& \frac{9}{2}(1-2K_i^2)(1-2K_j^2) \, \biggr] \biggr), \nonumber \\
K_2&=& \mu_2^*+\frac{1}{2}(\mu_2+\mu_1^*+\mu_3^*) - \frac{k(m+n)}{\sqrt{2}}, \nonumber \\
K_4&=& \frac{1}{2}(\mu_1^*+\mu_3^*-\mu_2)+\frac{k(m+n)}{\sqrt{2}}, \nonumber 
\end{eqnarray}
and 
\begin{eqnarray}
K_{2j-1} = \mu_{2j-1} + \frac{1}{2}(\mu_2+(-1)^j [\mu_3^*-\mu_1^* + \sqrt{2} k(m-n) ]) \nonumber 
\end{eqnarray}
($j=1$, $2$). As is shown in Fig.2, even the contribution of the O(${\rm tan}^2\chi$) term is quantitatively negligible, and, for this reason, higher order terms in ${\rm tan}\chi$ were neglected above. The mixing or coupling, occurring through the $k$-dependent terms in $p_0$ and $K_j$, between the momenta ${\bf p}$ on FS appearing in the gap function ${\hat \Delta}_p$ and the reciprocal lattice vectors leads to structural changes of the vortex lattice at a fixed orientation. By determining the $k$-value minimizing ${\cal F}_4$ ($>0$) at each ($H$, $T$), structural changes of the vortex lattice have been examined. 
\begin{figure}[t]
\scalebox{0.35}[0.35]{\includegraphics{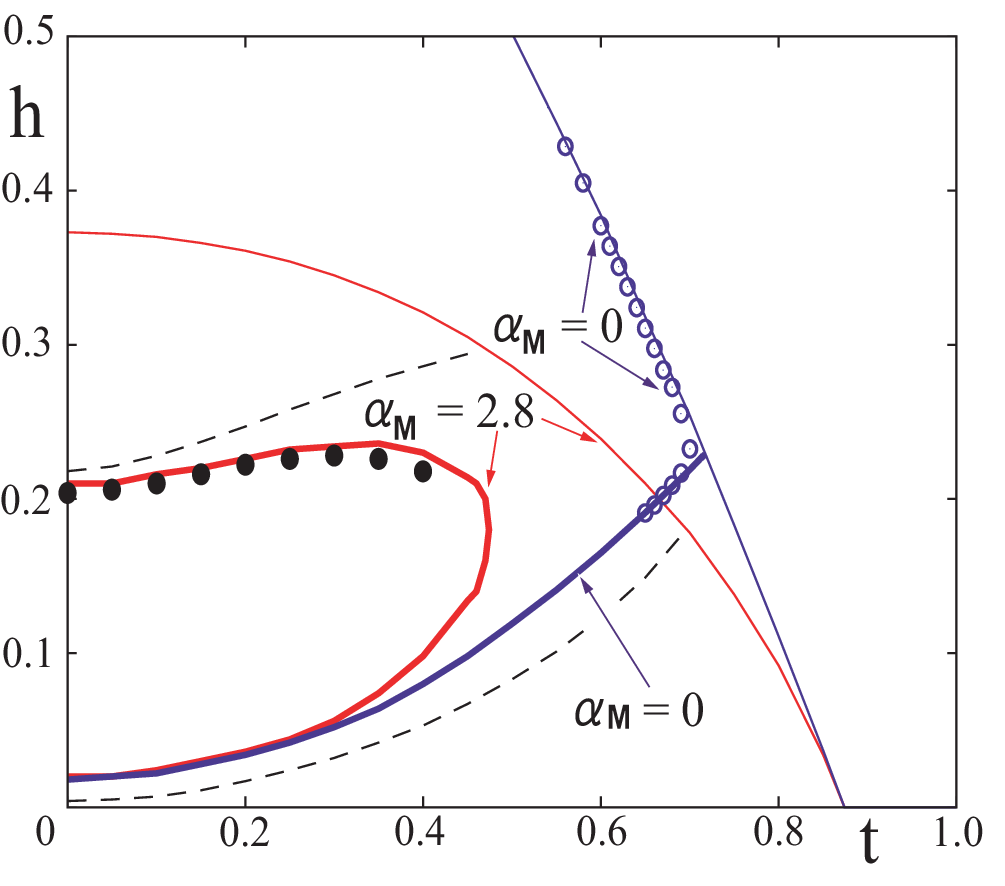}}
\caption{Square to rhombic ST (thick solid) curves and the corresponding $H_{c2}(T)$ (thin solid curves) in the $h$-$t$ phase diagram in the case with the Maki parameter $0$ and $2.8$, the circular FS, and the fixed value $l=20 \xi_0$. Each thick solid curve was obtained by neglecting the O(${\rm tan}^2\chi$) term in eq.(9). For $\alpha_M=0$, the square lattice is realized everywhere above the ST curve, while it is limited for $\alpha_M=2.8$ in the area surrounded by the closed ST curve. For $\alpha_M=2.8$, a {\it crossover} line on which $\gamma=70$ degrees and the high $H$ branch of the ST curve calculated by including the O(${\rm tan}^2\chi$) contribution of eq.(9) are indicated by the dashed curves and the solid circles, respectively. The open circles represent the $\alpha_M=0$ ST curve with fluctuation corrections (see the text). 
} \label{fig.2}
\end{figure}

As examples of calculation results following from eq.(9), we focus below on those for $\alpha_M =0$ and $2.8$. The resulting square to rhombic ST lines for these $\alpha_M$ values are expressed in Fig.2 by thick solid curves under a fixed $l/\xi_0$, where $t=T/T_{c0}$, and $h=H/H_{\rm orb}(0)$. In our calculation, the square to rhombic ST illustrated in Fig.1(a) was of second order everywhere. In higher $H$ where the paramagnetic depairing is more important, $\alpha_M$-dependence of the ST curve is striking: In the orbital limit where $\alpha_M=0$, the square lattice becomes more rigid in 
higher $H$, while it is limited, as in Fig.2, in the intermediate field range surrounded by a {\it closed} ST curve if $\alpha_M > 2.5$. This reentry of the rhombic lattice implies that the interplay between the orbital-depairing and the $d$-wave pairing symmetry, enhancing the square lattice and fixing the orientation of the vortex lattice, is weakened by the paramagnetic depairing. On the other hand, although the ST curve in low fields $h \ll 1$, where the paramagnetic depairing is ineffective, shows the expected behavior insensitive to $\alpha_M$-values, it is not quantitatively reliable because our approach assuming dominant roles of the lowest LL is valid in higher fields. In fact, the square lattice region seems to have been overestimated near the low $T$ and low $H$ corner. 

\begin{figure}[t]
\scalebox{0.35}[0.35]{\includegraphics{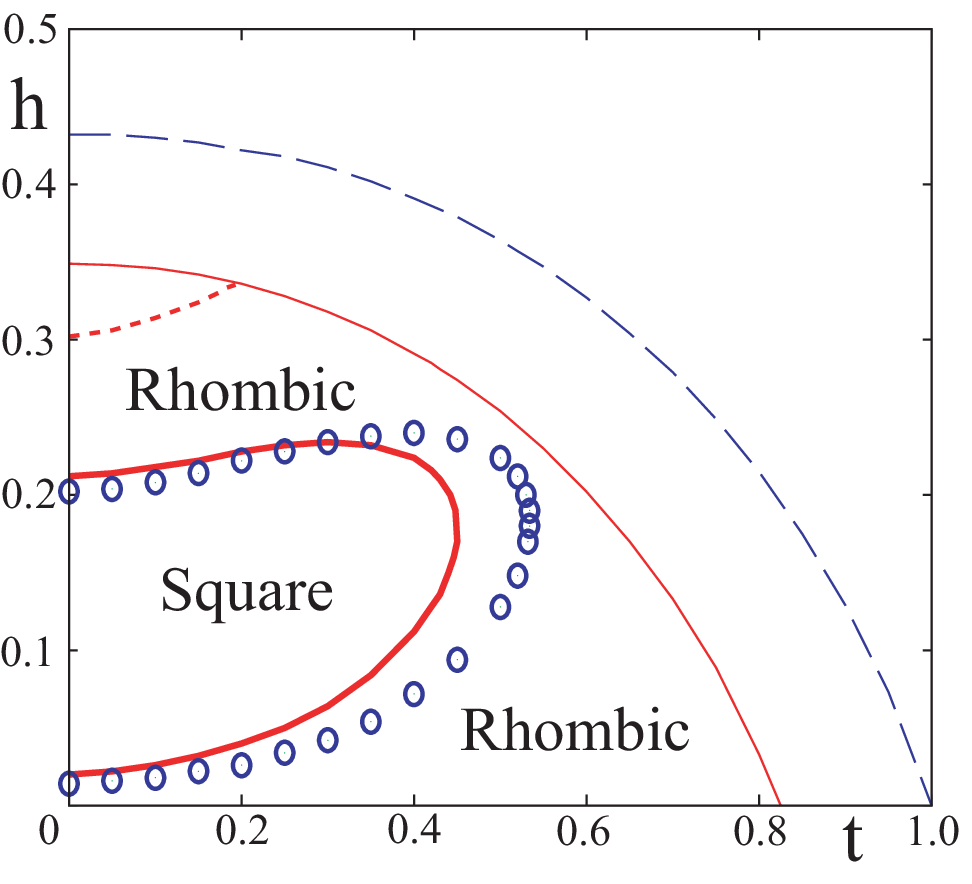}}
\caption{ST lines and the corresponding $H_{c2}(T)$ lines for the fixed value $\alpha_M=2.8$ obtained for $l=\infty$ (open circles and the upper dashed curve) and $l=14.5 \xi_0$ (thick and thin solid curves). When $l=\infty$, the {\it discontinuous} $H_{c2}$-transition \cite{AI} due to the paramagnetic depairing occurs in $t \leq 0.13$, and the Pauli-limiting field $H_P$ corresponds to the value $h=0.505$. Between the {\it first order} ST curve (lower dashed line) and $H_{c2}(T)$ in $\alpha_M=2.8$, the vortex lattice has the orientation indicated in Fig.1(b) \cite{Bianchi08}. } \label{fig.3}
\end{figure}

Figure 2 implies that a {\it reentrant} and closed ST curve similar to that \cite{Bianchi08} found from neutron scattering data of CeCoIn$_5$ in ${\bf H} \parallel c$ follows from $\alpha_M$ of order unity. Judging from such a large effect of a finite $\alpha_M$, the paramagnetic depairing is expected to be the main origin of the reentrant ST curve in CeCoIn$_5$ \cite{Bianchi08}. For comparison, an $\alpha_M=0$ ST curve with effects of elastic thermal fluctuation \cite{Kogan,Dorsey} included is expressed in Fig.2 by open circles. The fluctuation is incorporated in the squashing elastic modulus following from eq.(9) via replacement $k^2 \to k^2(1 - 3 k^2 {\overline {u^2}}/(4 \pi r_H^2))$ \cite{Dorsey}, where ${\overline {u^2}}$ is the mean square average of vortex displacement calculated in terms of the material parameters of CeCoIn$_5$ \cite{MS}. The obtained result clearly shows that the fluctuation-induced mechanism is a minor contribution to the reentry of the rhombic lattice in CeCoIn$_5$. 
Hereafter, two additional features seen in the experimental phase diagram will be discussed. First, the high $H$ branch of the ST curve in Ref.\cite{Bianchi08} has shown a negative slope even at lower $T$ in contrast to the positive slope in Fig.2. Since the paramagnetic effect suppressing the square lattice is more effective at lower $T$, the positive slope in the present calculation is reasonable. A possible origin changing the slope sign of the high $H$ branch is the quasiparticle damping $\xi_0/l$ due to the non-SC critical fluctuation. In Fig.3, 
we show $l/\xi_0$ dependence of the phase diagram in $\alpha_M=2.8$ case. In the context of CeCoIn$_5$, the quasiparticle mean free path $l$ is not due to impurity scatterings but rather should be a consequence of non-SC critical fluctuations. It has been argued \cite{RI07} that this damping effect is {\it not} negligible in the high $H$ region close to $H_{c2}(0)$ of CeCoIn$_5$ in ${\bf H} \parallel c$. As Fig.3 shows, effects of a nonvanishing $\xi_0/l$ on the high $H$ ST curve are quantitatively weak in agreement with our view that a large $\alpha_M$ is the main origin of the reentrant ST curve. Nevertheless, the high $H$ ST curve tends to approach a flat curve with decreasing $l/\xi_0$: In the low $T$ region dominated by the paramagnetic depairing, a quasiparticle damping suppressing the paramagnetic effect is more effective and slightly shifts the high $H$ branch upwardly, while it shifts this branch downwardly at higher $T$ where the orbital depairing is rather important. We expect an inclusion of a (unknown) $T$-dependence of $l$ to resolve this issue more satisfactorily. 

The experimental phase diagram \cite{Bianchi08} also includes {\it first order} STs between the two orientations indicated in Fig.1(a) and (b) both above and below the field region of the square lattice. To understand this, we have also examined effects of a weak four-fold anisotropy of FS {\it competitive}, in orientation, with that of ${\hat \Delta}_p$ on the high $h$ region of the $\alpha_M=2.8$ phase diagram by introducing the anisotropy, as in Ref.\cite{Nakai}, through the replacement \cite{Nakai} on the Fermi velocity, $v_F \to v_F (1 - \beta {\rm cos}4\phi_p)/\sqrt{1 - \beta^2}$ ($\beta > 0$), and the density of states. When this $v_F$-anisotropy is more dominant, the vortex lattice begins to change with the fixed orientation of Fig.1(b). We find that, for a weak FS anisotropy with $\beta=0.05$, a {\it first order} ST between the two orientations indicated in Fig.1 appears on the lower dashed curve in Fig.3, above which the square lattice and a rhombic one with $\gamma=74.5$ degrees are nearly degenerate in energy with the orientation of Fig.1 (b). This result fixes our view on the high field side of the experimental phase diagram of CeCoIn$_5$ \cite{Bianchi08} and supports the picture that the closed and reentrant ST curve is due not to a FS anisotropy but to the $d_{x^2-y^2}$-pairing symmetry of CeCoIn$_5$. Actually, if the in-plane FS anisotropy relevant to the superconductivity of CeCoIn$_5$ is characterized as a {\it single} four-fold anisotropy, it is quite unreasonable to ascribe both of the reentrant ST curve and the first order STs to such a single FS anisotropy. Further, bearing the result in Ref.\cite{Nakai} in mind, the fact that another first order ST in lower fields, where the finite $\alpha_{\rm M}$ does not work, is limited to a narrow range ($< 0.5$ (T)) \cite{Bianchi08} is consistent with the weak FS anisotropy assumed here. We note that the states with the orientation of Fig.1(b) are limited to the range $h < 0.4$ {\it irrespective of} the $l/\xi_0$ value and, when $l=\infty$, are not realized near $H_{c2}(T)$ in contrast to the observation \cite{Bianchi08}. This supports the argument \cite{RI07} that the quasiparticle damping is not negligible in understanding the region near $H_{c2}$ of CeCoIn$_5$. 

In conclusion, the paramagnetic depairing easily destroys the square vortex lattice stabilized by a $d$-wave pairing and is believed to be the main origin of the reentrant square to rhombic structural transition curve found in CeCoIn$_5$ \cite{Bianchi08}. The present results indicate that the paramagnetic depairing plays unexpectedly crucial roles in the high $H$ vortex lattice structure and may be the main origin of other field-induced lattice structure transitions such as the square to rhombic one in TmNi$_2$B$_2$C with a four-fold anisotropic Fermi surface \cite{Eskildsen}. 

We are grateful to H. Adachi, Y. Matsuda, T. Shibauchi, and M. R. Eskildsen for useful discussions.

\end{document}